\documentclass[twocolumn]{svjour3}                     
\usepackage{graphicx}
\usepackage[round]{natbib}
\begin{document}

\title{Diffusive Nested Sampling\thanks{Software (in C++) implementing Diffusive Nested Sampling is available at \texttt{http://lindor.physics.ucsb.edu/DNest/} under the GNU General Public License.}
}


\author{Brendon J. Brewer        \and
        Livia B. P{\'a}rtay        \and
	G{\'a}bor Cs{\'a}nyi
}


\institute{B. J. Brewer \at
              Department of Physics, University of California, Santa Barbara, CA 93106-9530, USA \\
              \email{brewer@physics.ucsb.edu}           
           \and
           L. B. P{\'a}rtay \at
           University Chemical Laboratory, University of Cambridge, Lensfield Road, CB2 1EW, Cambridge, United Kingdom
           \and
	   G. Cs{\'a}nyi \at
	   Engineering Laboratory, University of Cambridge, Trumpington Street, CB2 1PZ Cambridge, United Kingdom
}

\date{}

\maketitle

\begin{abstract}
We introduce a general Monte Carlo method based on Nested Sampling (NS), for sampling complex probability distributions and estimating the normalising constant. The method uses one or more particles, which explore a mixture of nested probability distributions, each successive distribution occupying $\sim e^{-1}$ times the enclosed prior mass of the previous distribution. While NS technically requires independent generation of particles, Markov Chain Monte Carlo (MCMC) exploration fits naturally into this technique. We illustrate the new method on a test problem and find that it can achieve four times the accuracy of classic MCMC-based Nested Sampling, for the same computational effort; equivalent to a factor of 16 speedup. An additional benefit is that more samples and a more accurate evidence value can be obtained simply by continuing the run for longer, as in standard MCMC. 
\end{abstract}

\section{Introduction}
In probabilistic inference and statistical mechanics, it is often necessary to characterise complex multidimensional probability distributions. Typically, we have the ability to evaluate a function proportional to the probability or probability density at any given point. Given this ability, we would like to produce sample points from the target distribution (to characterise our uncertainties), and also evaluate the normalising constant. In Bayesian Inference, a prior distribution $\pi(\theta)$ is modulated by the likelihood function $L(\theta)$ to produce the posterior distribution $f(\theta)$:
\begin{equation}
f(\theta) = \frac{1}{Z} \pi(\theta) L(\theta)
\end{equation}
where $Z = \int \pi(\theta) L(\theta) \, d\theta$ is the evidence value, which allows the entire model to be tested against any proposed alternative. In statistical mechanics, the goal is to produce samples from a canonical distribution
\begin{equation}
f(\theta) = \frac{g(\theta)}{Z(\lambda)} \exp\left(-\lambda E(\theta)\right)
\end{equation}
($g(\theta) = $ density of states, $E(\theta) = $ energy) at a range of inverse-temperatures $\lambda$; and also to obtain the normalisation $Z(\lambda)$ as a function of $\lambda$. The challenge is to develop sampling algorithms that are of general applicability, and are capable of handling the following complications: multimodality, phase transitions, and strong correlations between variables (particularly when the Markov Chain Monte Carlo (MCMC) proposals are not aware of these correlations). These challenges have led to the development of a large number of sophisticated and efficient techniques \citep[e.g.][]{chib, veitch}. However, our goal in the present paper is to develop a method that requires little problem-specific tuning, obviating the need for large numbers of preliminary runs or analytical work - Diffusive Nested Sampling can be applied to any problem where Metropolis-Hastings can be used. Finally, we note that no sampling method can ever be completely immune from failure; for example, it is difficult to imagine any method that will find an exceedingly small, sharp peak whose domain of attraction is also very small.

\subsection{Nested Sampling}
Nested Sampling (NS) is a powerful and widely applicable algorithm for Bayesian computation \citep{skilling}. Starting with a population of particles $\{\theta_i\}$ drawn from the prior distribution $\pi(\theta)$, the worst particle (lowest likelihood $L(\theta)$) is recorded and then replaced with a new sample from the prior distribution, subject to the constraint that its likelihood must be higher than that of the point it is replacing. As this process is repeated, the population of points moves progressively higher in likelihood.

Each time the worst particle is recorded, it is assigned a value $X \in [0,1]$, which represents the amount of prior mass estimated to lie at a higher likelihood than that of the discarded point. Assigning $X$-values to points creates a mapping from the parameter space to $[0,1]$, where the prior becomes a uniform distribution over $[0,1]$ and the likelihood function is a decreasing function of $X$. Then, the evidence can be computed by simple numerical integration, and posterior weights can be assigned by assigning a width to each point, such that the posterior mass associated with the point is proportional to \texttt{width} $\times$ \texttt{likelihood}.

The key challenge in implementing Nested Sampling for real problems is to be able to generate the new particle from the prior, subject to the hard likelihood constraint. If the discarded point has likelihood $L^*$, the newly generated point should be sampled from the constrained distribution:
\begin{equation}\label{constrained}
p_{L^*}(\theta) = \frac{\pi(\theta)}{X^*}\left\{\begin{array}{lr}1, & L(\theta) > L^*\\ 0, & \textnormal{otherwise.}\end{array}\right.
\end{equation}
where $X^*$ is the normalising constant. Technically, our knowledge of this new point should be independent of all of the surviving points. A simple way to generate such a point is suggested by \citet{sivia}: copy one of the surviving points and evolve it via MCMC with respect to the prior distribution, rejecting proposals that would take the likelihood below the current cutoff value $L^*$. This evolves the particle with Equation~\ref{constrained} as the target distribution. If the MCMC is done for long enough, the new point will be effectively independent of the surviving population and will be distributed according to Equation~\ref{constrained}. Throughout this paper we refer to this strategy as ``classic'' Nested Sampling.

However, in complex problems, this approach can easily fail - constrained distributions can often be very difficult to efficiently explore via MCMC, particularly if the target distribution is multimodal or highly correlated. To overcome these drawbacks, several methods have been developed for generating the new particle \citep{2006ApJ...638L..51M, 2008arXiv0809.3437F} and these methods have been successful in improving the performance of Nested Sampling, at least in low-dimensional problems. Techniques for diagnosing multimodality have also been suggested by \citet{2009arXiv0906.3544P}. In section~\ref{multi} we introduce our multi-level method which retains the flexibility and generality of MCMC exploration, and is also capable of efficiently exploring difficult constrained distributions.

The main advantage of Nested Sampling is that successive constrained distributions (i.e. $p_{L^*_j}(\theta), p_{L^*_{j+1}}(\theta)$, and so on) are, by construction, all compressed by the same factor relative to their predecessors. This is not the case with tempered distributions of the form $p_T(\theta) \propto \pi(\theta)L(\theta)^{1/T}$, where a small change in temperature $T$ can correspond to a small or a large compression. Tempering based algorithms (e.g. simulated annealing, parallel tempering) will fail unless the density of temperature levels is adjusted according to the specific heat, which becomes difficult at a first-order phase transition \citep[the uniform-energy sampling of][is also incapable of handling first-order phase transitions]{wl}. Unfortunately, knowing the appropriate values for the temperature levels is equivalent to having already solved the problem. Nested Sampling does not suffer from this issue because it asks the question ``what should the next distribution be, such that it will be compressed by the desired factor'', rather than the tempering question ``the next distribution is pre-defined, how compressed is it relative to the current distribution?''

\section{Multi-Level Exploration}\label{multi}
Our algorithm begins by generating a point from the prior ($\pi(\theta) = p_{L^*_0}$ where $L^*_0 = 0$), and evolving it via MCMC (or independent sampling), storing {\it all} of the likelihood values of the visited points. After some predetermined number of iterations, we find the $1-e^{-1} \approx 0.63212$ quantile of the accumulated likelihoods, and record it as $L^*_1$, creating a new level that occupies about $e^{-1}$ times as much prior mass as $p_{L^*_0}$. Likelihood values less than $L^*_1$ are then removed from the accumulated likelihood array.

Next, classic Nested Sampling would attempt to sample $p_{L^*_1}$ via MCMC. The difference here is that we attempt to sample a {\it mixture} (weighted sum) of $p_{L^*_0}$ and $p_{L^*_1}$. Thus, there is some chance of the particle escaping to lower constrained distributions, where it is allowed to explore more freely. Once we have enough samples from $p_{L^*_1}$ (equivalent to all points from the mixture that happen to exceed $L^*_1$), we find the $1-e^{-1} \approx 0.63212$ quantile of these likelihoods and record it as $L^*_2$. Likelihood values less than $L^*_2$ are then removed from the accumulated likelihood array. The particle then explores a mixture of $p_{L^*_0}$, $p_{L^*_1}$ and $p_{L^*_2}$, and so on.

Each time we create a new level, its corresponding constrained distribution covers about $e^{-1}$ as much prior mass as the last. Thus, we can estimate the $X$-value of the $k$th level created as being $\exp\left(-k\right)$. This estimate can be refined later on, as explained in Section~\ref{revising}.

Once we have obtained the desired number of levels, we allow the particle to continue exploring a mixture of all levels. This multi-level exploration scheme is similar to simulated tempering \citep{1992EL.....19..451M}, but using the well-tuned $L^*$ values from the run, rather than using a pre-defined sequence of temperatures that may be poorly adapted to the problem at hand.

\subsection{Weighting Schemes}\label{weighting}
The use of a mixture of constrained distributions raises the question: how should we weight each of the mixture components? Naive uniform weighting works, but takes time proportional to $N^2$ to create $N$ levels; contrast this with classic Nested Sampling which has $O(N)$ performance in this regard. A simple parametric family of weighting schemes are the exponentially-decaying weights:
\begin{equation}
\begin{array}{lr}
w_j \propto \exp \left(\frac{j - J}{\Lambda}\right) & , \textnormal{for } j \in \{1, 2, ..., J\}
\end{array}
\end{equation}
where $J$ is the current highest level, and $\Lambda$ is a scale length, describing how far we are willing to let the particle ``backtrack'' to freer distributions, in order to assist with exploration and the creation of a new, higher level. With this exponential choice, the time taken to create $N$ levels is proportional to $N$, the proportionality constant being dependent on $\Lambda$. Smaller values are more aggressive, and larger values, while slower, are more fail-safe because they allow the particle to explore for longer, and more freely.

Once the desired number of levels has been created, the weights $\{w_j\}$ are changed to uniform, to allow further samples to be drawn. Non-uniform weights can also be used, for example to spend more time at levels that are significant for the posterior. The simulation can be run for as long as required, and the evidence and posterior samples will converge in a manner analogous to standard MCMC. Additional seperate runs are not required, provided that enough levels have been created.

\begin{figure*}
\begin{center}
\includegraphics[scale=0.6]{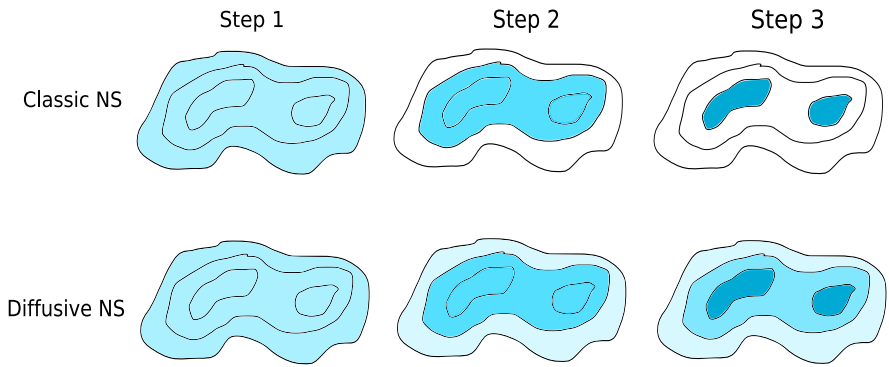}
\caption{An illustration of a the distributions that must be sampled as Nested Sampling progresses. In the classic scheme, at Step 3, we must obtain a sample from the coloured region which is composed of two separate islands, which is usually very difficult if MCMC is the only exploration option. To ameliorate these difficulties, we explore the mixture distribution (bottom right), where travel between isolated modes is more likely.\label{nested}}
\end{center}
\end{figure*}

\subsection{Exploring the mixture}
Suppose we have obtained a set of constrained distributions $p_{L^*_j}(\theta)$ from the algorithm described above (Section~\ref{multi}). Each of the constrained distributions is defined as follows:
\begin{equation}
p_{L^*_j}(\theta) = \frac{\pi(\theta)}{X_j}\left\{\begin{array}{lr}1, & L(\theta) > L^*_j\\ 0, & \textnormal{otherwise.}\end{array}\right.
\end{equation}
The particle in our simulation, $\theta$, is assigned a label $j$ indicating which particular constrained distribution it is currently representing. $j=0$ denotes the prior, and $j=1,2,3,...,  J$ denote progressively higher likelihood cutoffs, where $j=J$ is the current top level. To allow $\theta$ to explore the mixture, we update $\left\{\theta, j\right\}$ so as to explore the joint distribution
\begin{eqnarray}
p(\theta, j) &=& p(j) p(\theta|j) \nonumber \\
&=& \frac{w_j}{X_j} \pi(\theta)\left\{\begin{array}{lr}1, & L(\theta) > L^*_j\\ 0, & \textnormal{otherwise.}\end{array}\right.\label{mixture}
\end{eqnarray}
where $p(j) = w_j$ is the chosen weighting scheme discussed in Section~\ref{weighting}, and $X_j$ is the normalising constant for $p(\theta|j)$, which in general depends on $j$. In fact, $X_j$ is simply the amount of prior mass enclosed at a likelihood above $L^*_j$, and is the abcissa in the standard presentation of Nested Sampling \citep{skilling}, hence the notation $X$ rather than the usual $Z$ for a normalising constant.

Updating $\theta$ is done by proposing a change that keeps the prior $\pi(\theta)$ invariant, and then accepting as long as the likelihood exceeds $L^*_j$. To update $j$, we propose a new value from some proposal distribution (e.g., move either up or down one level, with 50\% probability) and accept using the usual Metropolis rule, with Equation~\ref{mixture} as the target distribution. This cannot be done without knowing the $\{X_j\}$ - however, our procedure for adding new levels ensures that the ratio of the $X$'s for neighbouring levels is approximately $e^{-1}$. So we at least have some approximate values for the $\{X_j\}$, and these can be used when updating $j$. However, since our creation of new levels is based on sampling, and therefore not exact, the ratios of $X$'s will be a little different from the theoretical expected value $e^{-1}$. Hence, we will actually be exploring Equation~\ref{mixture} with a marginal $p(j)$ that is slightly different than the weights $\{w_j\}$ that we really wanted. However, we can further revise our estimates of the $X$'s at each step to achieve more correct exploration; our method for doing this is explained in Section~\ref{revising}. This refinement not only allows the particle to explore the levels with the desired weighting, but also increases the accuracy of the resulting evidence estimate.

Note that it is possible to explicitly marginalise over $j$ and explore a distribution for $\theta$ only, however it is convenient to retain $j$ for various purposes, most notably the refinements in Sections~\ref{revising} and~\ref{stuck}.

\section{Revising the $X$'s}\label{revising}
As sampling progresses and more levels are added, the actual $X$-values of the levels can become different from the theoretical expectation $X_{j+1}=e^{-1}X_j$ which would be realised if our sampling were perfect. This causes the simulation to explore a $p(j)$ that is very different from the desired weights $\{w_j\}$. Fortunately, we can use details of the exploration to obtain estimates of the $X$'s that are more accurate than those given by the theoretical expectation $X_{j+1}=e^{-1}X_j$. Conditioning on a particular level $j$, the particle's likelihood should exceed level $j+1$'s likelihood cutoff a fraction $X_{j+1}/X_j$ of the time. Thus, we can use the {\it actual} fraction of exceedences $n(L>L^*_{j+1}|j)/n(j)$ as an estimate of the true ratio of normalising constants for consecutive levels. In practice, we only keep track of the exceedence fraction for consecutive levels, and use the theoretical expected value $e^{-1}$ to stabilise the estimate when the number of visits $n(j)$ is low:
\begin{equation}\label{estimate}
\frac{X_{j+1}}{X_{j}} \approx \frac{n(L>L^*_{j+1}|j) + Ce^{-1}}{n(j) + C}
\end{equation}
Here, the number $C$ represents our effective confidence in the theoretical expected value relative to the empirical estimate, and thus resembles a Bayesian estimate. The theoretical estimate $e^{-1}$ dominates the estimate until the sample size $n(j)$ exceeds $C$, whereupon the empirical estimate becomes more important. Clearly, $n(j)$ should not be accumulated until level $j+1$ exists. An additional refinement of this approach can be made by realising that the particle, even if it is at level $j$, may also be considered as sampling higher distributions if its likelihood happens to exceed the higher levels' thresholds.

\subsection{Enforcing the Target Weighting}\label{stuck}
Note that the ratio of normalising constants between level $j$ and level $j+1$ is estimated using only samples at level $j$. Sometimes, if the upper levels have been created incorrectly (typically they are too closely spaced in $\log X$), the particle spends too much time in the upper levels, rarely spending time in the lower levels, which would be needed to correct the erroneous $X$-values.

To work around this issue, we also keep track of the number of visits to each level relative to the expected number of visits, and add a term to the acceptance probability for moving $j$, to favour levels that haven't been visited as often as they should have. If a move is proposed from level $j$ to level $k$, the Metropolis-Hastings acceptance probability is multiplied by the following factor:
\begin{equation}
\alpha_{{\rm enforce}} = \left(\frac{(n_j + C)/(\left<n_j\right> + C)}{(n_k + C)/(\left<n_k\right> + C)}\right)^\beta
\end{equation}
where $n_j$ and $n_k$ are the number of visits to levels $j$ and $k$ respectively (These are different from $n(j)$ in Section~\ref{revising}, as they can be accumulated immediately, not needing a higher level to exist). Thus, the transition is favoured if it moves to a level $k$ that has not been visited as often as it should have ($\left<n_k\right>$). This procedure is analogous to the approach used by \citet{wl} to sample a uniform distribution of energies. These expected numbers of visits are tracked throughout the entire simulation, they are essentially the integrated values of the weights $\{w_j\}$ over the history of the simulation. The power $\beta$ controls the strength of the effect, and $C$ again acts to regularise the effect when the number counts are low.

This procedure (and the one in Section~\ref{revising}) destroys the Markov property, but this effect decreases towards zero as the simulation proceeds, so the eventual convergence of the algorithm is not affected \citep[this is analogous to the kind of tuning discussed by][]{roberts, rosenthal}. In practice, any biases introduced by this procedure (even with a strong choice of $\beta=10$) have been found to be negligible relative to the dominant source of error, which is incorrect estimates of the $X$'s of the levels.

\section{Assigning X-values to samples}
A diffusive nested sampling run explores the joint distribution of Equation~\ref{mixture} in an MCMC-style way. Usually this involves a lot of steps, and it is wasteful (of disk space and I/O overhead) to save them all; therefore it is necessary to ``thin the chain''. This results in an output sample of $\theta$ points. To use them for calculating the evidence and posterior quantities, these points must be assigned $X$-values.

Firstly, each $\theta$ in the sample is assigned an {\it interval} of possible $X$-values, by finding the two levels whose likelihoods sandwich the particle. The conditional distribution of the $X$-values of points given that they lie in some interval is uniform, so we can assign $X$-values by generating uniform random variates between the $X$-values of the two sandwiching levels. This probabilistic assignment of $X$-values to the samples gives error bars on the evidence and posterior quantities, as classic NS does. However, it assumes that the $X$-values of the levels are known exactly, which they are not. Unfortunately, this error tends to be more significant than the uncertainty of not knowing where each particle lies within its interval, and the error bars are over-optimistic as a result. Unfortunately, obtaining reliable error bars from MCMC-based computation remains difficult. Possible methods for assigning uncertainties to the $X$-values of the {\it levels} would make use of the number counts $n(L>L^*_{j+1}|j)$, however such a scheme has not been implemented at the time of writing.

\section{Test Problem}\label{test}
In this section we demonstrate the diffusive Nested Sampling method on a test problem that is quite simple, yet has enough complications to shed some light on the differences between the diffusive and classic NS algorithms. The problem is adapted from \citet{skilling}'s ``Statistics Problem'' but modified to be bimodal. The parameter space is 20-dimensional, and the prior is uniform between $[-0.5, 0.5]$ in each dimension:
\begin{equation}
\pi(x_1, x_2, ..., x_{20}) = \prod_{i=1}^{20} \left\{\begin{array}{lr} 1, & x_i \in [-0.5, 0.5] \\ 0, & \textnormal{otherwise}\end{array}\right.
\end{equation}
The likelihood function is the sum of two Gaussians, one centred at the origin and having width $v=0.1$ in each dimension, and the other centred at $(0.031, 0.031, ..., 0.031)$ and having width $u=0.01$. The narrower peak is weighted so that it contains 100 times as much posterior mass as the broad peak. The likelihood function is:
\begin{eqnarray}
L(x_1, x_2, ..., x_{20}) = \prod_{i=1}^{20}\frac{\exp\left(-\frac{1}{2}(x_i/v)^2\right)}{v\sqrt{2\pi}} \\ + 100\prod_{i=1}^{20}\frac{\exp\left(-\frac{1}{2}((x_i - 0.031)/u)^2\right)}{u\sqrt{2\pi}}
\end{eqnarray}
For this problem the true value of the evidence is $\log(101)$ $\approx 4.6151$, to a very good approximation. \citet{skilling} showed that this system has a phase transition which is handled easily by classic Nested Sampling. By shifting the dominant mode off-centre, now the problem has a phase transition and is also bimodal. The bimodality allows us to illustrate an interesting effect where imperfect MCMC exploration causes the dominant mode to be discovered late (Section~\ref{typical}).

The MCMC transitions used throughout this section were all naïve random walk transitions, centred on the current value, with one of the $x$'s chosen at random to be updated. To work around the fact that the optimal step-size changes throughout the run, we draw the step size $S$ randomly from a scale-invariant Jeffreys prior $\propto 1/S$ at each step, with an upper limit for $S$ of $10^{0} = 1$ and a lower limit of $10^{-6}$. The proposal distribution used to change $j$ was a Gaussian centred at the current value, with standard deviation $S'$ which was chosen from a Jeffreys prior between 1 and 100 at each step. The proposed $j$ was then, of course, rounded to an integer value and rejected immediately if it fell outside of the allowed range of existing levels. A more sophisticated approach to setting the stepsize would be to use Slice Sampling \citep{slice}.

\subsection{Typical Performance of Diffusive NS}\label{typical}
\begin{figure*}
\begin{center}
\includegraphics[scale=0.5]{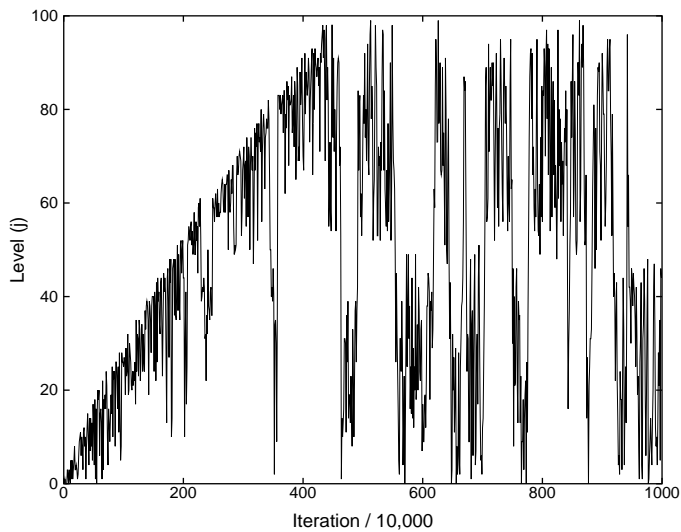}
\caption{The anatomy of a diffusive-NS run. Initially, the particle tends to move upwards, creating new likelihood levels. During this phase, the particle can backtrack somewhat, allowing freer exploration and also refining the estimates of the compression ratios of the newly created levels. Once the desired number of levels (in this case, 100) have been created, the particle attempts to explore all levels with the prescribed weights, in this case uniform weights.\label{explore}}
\end{center}
\end{figure*}

To illustrate the typical output from a diffusive Nested Sampling run, we executed the program on the test problem and using numerical parameters defined in Table~\ref{params}. The progress of the algorithm is displayed graphically in Figure~\ref{explore}. Diffusive NS has two distinct stages, the initial mostly-upward progress of the particle, and then an exploration stage (in this case, exploring all levels uniformly) which can be run indefinitely, and is constantly generating new samples and refining the estimates of the level $X$-values.

\begin{figure*}
\begin{center}
\includegraphics[scale=0.5]{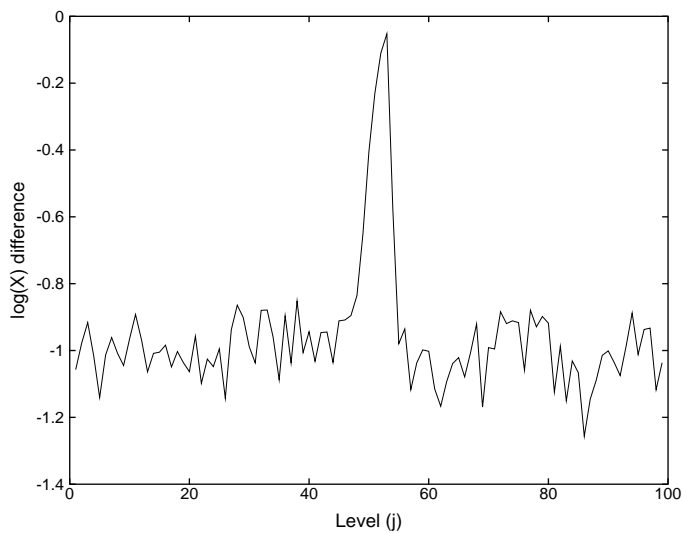}
\caption{Estimated compression from each level's distribution to the next. If sampling were perfect, the $\log(X)$ difference between successive levels would be $-1$. This is approximately correct except around levels 50-55, where some levels were ``accidentally'' created too close together. This is because the slowly-exploring particle failed to notice the presence of the narrow peak immediately, but in the meantime still created some new levels. \label{level_differences}}
\end{center}
\end{figure*}

\begin{figure*}
\begin{center}
\includegraphics[scale=0.6]{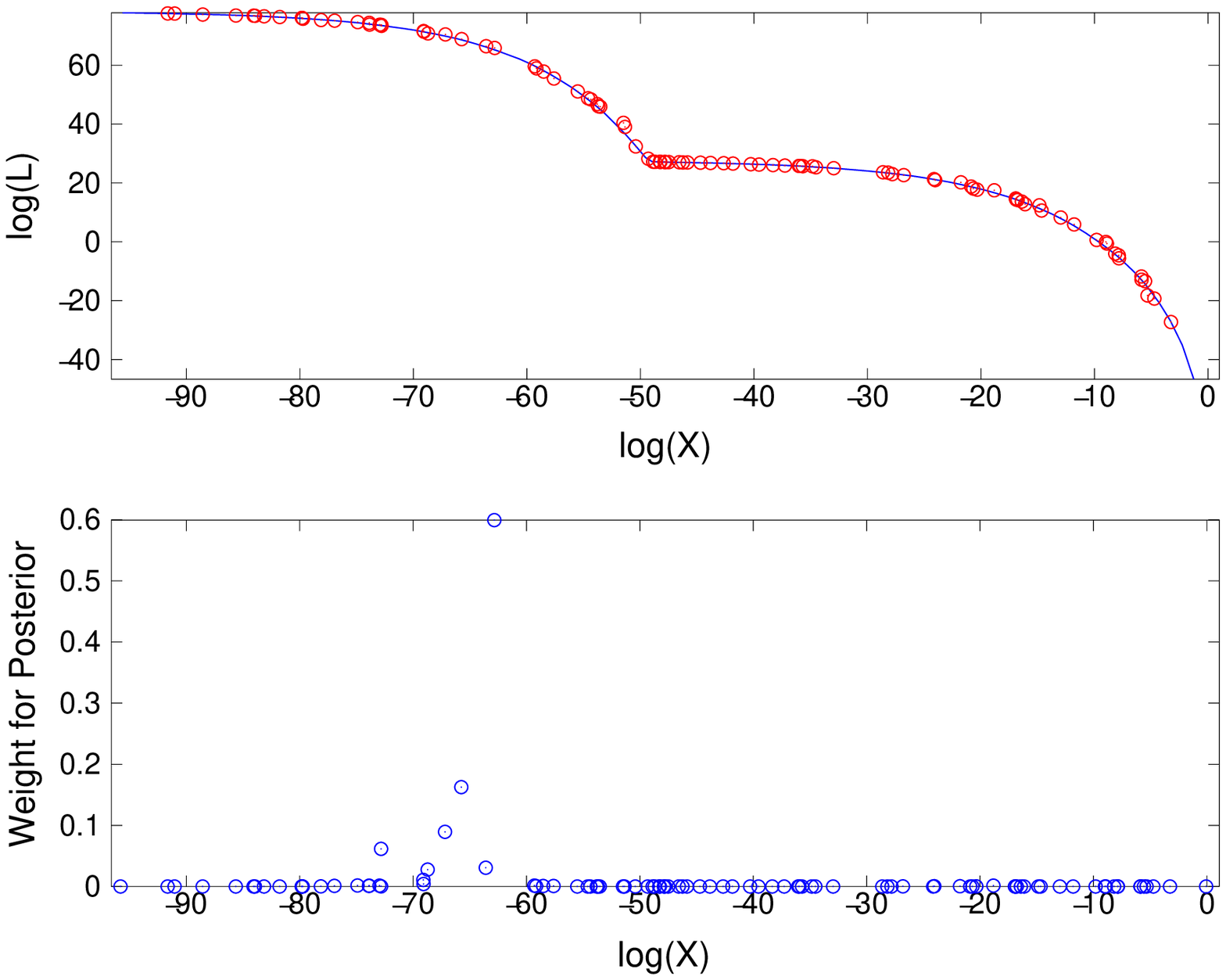}
\caption{Estimated log-likelihood curve for the test problem, based on a diffusive Nested Sampling run. The blue curve is constructed from the estimated $X$-values of the levels, while the circles represent sample points. The chain has been thinned further in order to show the discrete points more clearly.\label{logl_curve}}
\end{center}
\end{figure*}

Of particular note is the fact that, during the uniform exploration phase, the particle can easily mix between levels 55 and 100, and between 0 and 55, but jumps between these regions occur more rarely. This occurs around where the narrow peak in the posterior becomes important; imperfect MCMC exploration does not notice its presence right away.See Figure~\ref{level_differences} for more information about this phenomenon. The final log-likelihood vs prior-mass curve, and the output samples, are shown in Figure~\ref{logl_curve}.

\begin{table}
\begin{center}
\caption{Parameter values used for diffusive nested sampling on the test problem.\label{params}}
\begin{tabular}{ll}
\hline
Parameter & Value \\
\hline
Number of particles & 1\\
Number of likelihoods needed to create new level & 10,000 \\
Interval between particle saves & 10,000 \\
Maximum number of levels & 100 \\
Regularisation constant ($C$) & 1,000 \\
Degree of backtracking ($\Lambda$) & 10 \\
Strength of effect to enforce exploration weights ($\beta$) & 10 
\end{tabular}
\end{center}
\end{table}

\subsection{Empirical comparisons with Classic NS}
It is of interest to evaluate the performance of the diffusive NS method in comparison to classic NS. Is it less, equally, or more efficient? In particular, how well do the two methods cope with imperfect MCMC exploration?

In this section we explore this issue, albeit in a limited way. We ran diffusive and classic NS on the test problem defined in Section~\ref{test}, with the parameters of the algorithms chosen to be as similar. Both algorithms were limited to $10^7$ likelihood evaluations for all of the tests. The parameters for the diffusive NS runs are shown in Table~\ref{params}, and the classic NS parameters were chosen such that the programs have reached $\log X = -100$ by the time the alloted $10^7$ likelihood evaluations had occurred. These parameters are listed in Table~\ref{empirical}. The number of steps parameter listed in Table~\ref{empirical} for classic NS defines how many MCMC steps were used in an attempt to equilibrate a particle. The implementation of the classic Nested Sampling algorithm was the standard one where a particle is copied before being evolved.

\begin{figure*}
\begin{center}
\includegraphics[scale=0.4]{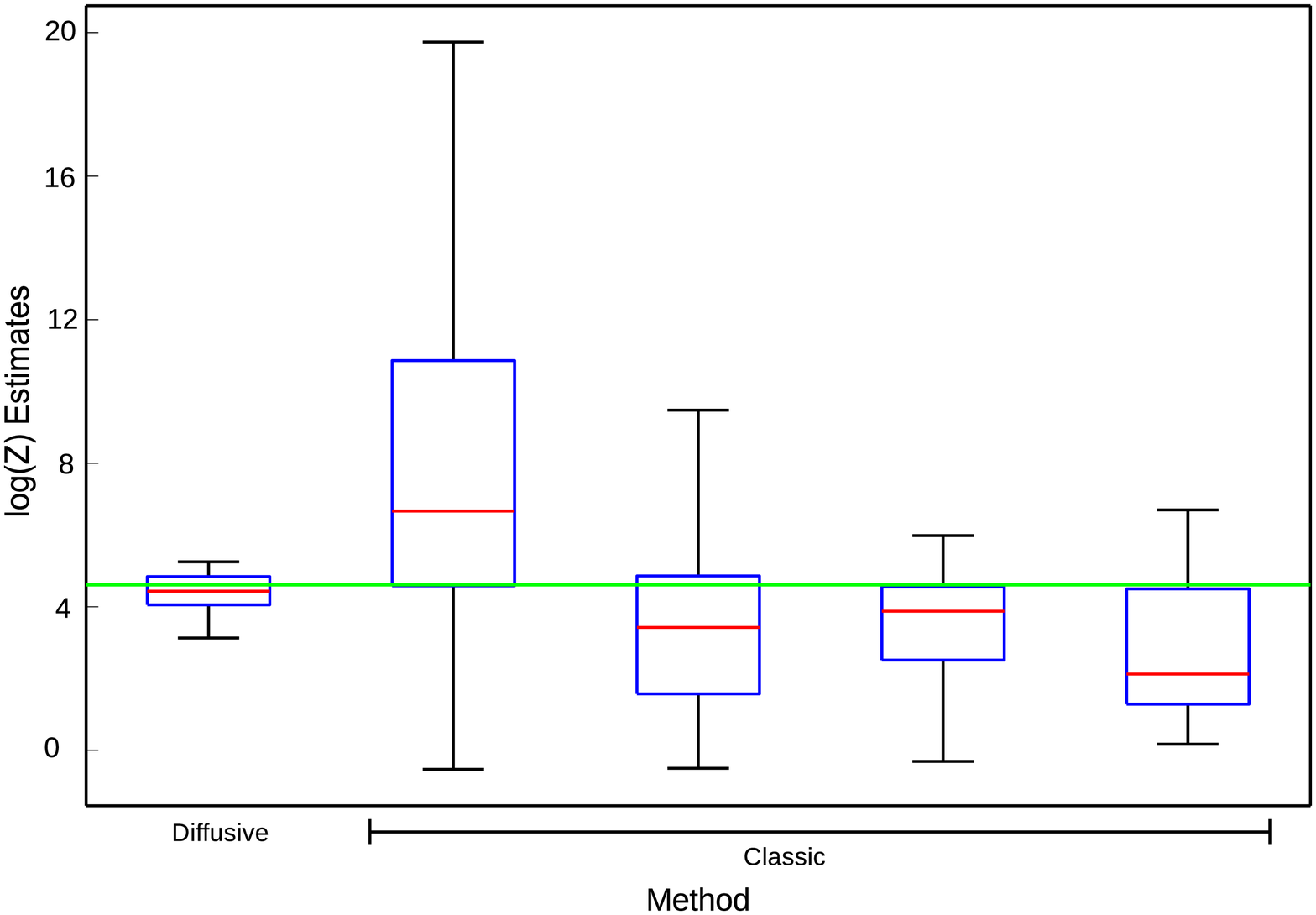}
\caption{Performance of five different methods on the test problem. The first method (leftmost box and whisker plot) is diffusive NS, the others are classic NS in increasing order of the number of particles. The bars indicate the spread of results obtained from repeated runs of each method with different random number seeds. The horizontal line indicates the true value.\label{methods}}
\end{center}
\end{figure*}

Each of these algorithms were run 24 times, and the root mean square error of the estimated (posterior mean) evidence values were calculated. These results are shown in Table~\ref{empirical} and Figure~\ref{methods}, and show a significant advantage to diffusive NS, by a factor of about four even compared to the most favourable classic NS parameters. To obtain similar accuracy with classic NS would have required 16 times as much computation.

This improvement can be attributed to diffusive NS's use of all visited likelihoods when creating a new level. Classic NS uses only the likelihood of the endpoint of the equilibration process, yet presumably all of the likelihoods encountered during the exploration are relevant for creating a new level, or for judging the compression of an existing level.

Another feature of diffusive NS that probably contributes to its efficiency is the fact that the particle can backtrack and explore with respect to broader distributions. This is particularly important when the target distribution is correlated, yet the proposal distribution is not. Then, the particle can fall down several levels, take larger steps across to the other side of the correlated target distribution, and then climb back up to the target distribution.

For classic NS with a single particle, the RMS error is comparable with the theoretical $\sqrt{H/N} \approx \sqrt{63.2/N}$ error bar of classic NS. The RMS error decreases with more particles, as expected \citep{murray}, but fails to fall off as $1/\sqrt{N}$, eventually increasing again. This is because the number of MCMC steps per iteration had to be decreased to compensate for the computational overhead of the increase in the number of particles, and the theoretical error bars assume that the exploration is perfect. While it may be possible to improve classic NS by attempting to retain the diversity of the initial population (sometimes evolving the deleted particle, rather than copying), this has not been implemented yet, and thus the comparison presented in this paper is against the common implementation of classic NS.

\begin{table*}
\begin{center}
\caption{The parameters for the test runs of the algorithms, and the resulting RMS error (from 24 runs of each algorithm) of the log evidence. Each algorithm was allowed $10^7$ likelihood evaluations. Diffusive NS outperformed classic NS significantly on this problem.\label{empirical}}
\begin{tabular}{llcc}
\hline
Algorithm & Parameter Values & RMS Deviation in log$(Z)$ & Theoretical $\sqrt{H/N}$\\
\hline
Diffusive NS & As per Table~\ref{params} & 0.583 & -\\
Classic NS & 1 particle, 100,000 MCMC steps per NS step& 5.82 & 7.95\\
Classic NS & 10 particles, 10,000 MCMC steps per NS step& 2.96 & 2.51\\
Classic NS & 100 particles, 1,000 MCMC steps per NS step & 2.07 & 0.80\\
Classic NS & 300 particles, 333 MCMC steps per NS step & 2.71 & 0.46
\end{tabular}
\end{center}
\end{table*}

\subsection{Multi-Particle Diffusive NS}
On highly correlated and multimodal target distributions, two strategies suggest themselves. One is to make $\Lambda$ large, so that the particle can backtrack a long way. Then, when the particle reaches the top level again, it is likely to be very weakly correlated with the original point. A more efficient alternative would be to run diffusive NS with multiple particles, evolving them in succession or choosing one at random at each step. If a particle gets stuck in a subordinate mode, this can be detected by its repeated failure to be within a distance of a few times $\Lambda$ of the top level $J$. Such particles can simply be deleted, and more CPU time can be used for evolving the surviving particles. This approach performs better than the copying method of classic NS because particles are only deleted when they are known to be failing, whereas in classic NS the copying operation will inevitably destroy the diversity of the initial population even when the MCMC exploration is satisfactory.

When running multi-particle diffusive NS, it is advisable to reset the counters (from Sections~\ref{revising} and~\ref{stuck}) once the desired number of levels have been created. This prevents the deleted particles from creating unnecessary barriers to exploration.

\section{Conclusions}
In this paper we introduced a variant of Nested Sampling that uses MCMC to explore a mixture of progressively constrained distributions. This method shares the advantages of the classic Nested Sampling method, but has several unique features. Firstly, once the desired number of constrained distributions have been created, the particle is allowed to diffusively explore all levels, perpetually obtaining more posterior samples and refining the estimate of the evidence. Secondly, this method uses the entire exploration history of the particle in order to estimate the enclosed prior mass associated with each level, and hence tends to estimate the enclosed prior mass of each level more accurately than the classic Nested Sampling method. We ran simple tests of the new algorithm on a test problem, and found that diffusive Nested sampling gives more accurate evidence estimates for the same computational effort. Whether this will hold true in general, or is a problem-dependent advantage, will be explored in the future.

\section{Acknowledgements}
We would like to thank John Skilling and Iain Murray for helpful discussions and comments on earlier versions of this paper. This work grew directly out of discussions that took place at MaxEnt 2009.


\begin{thebibliography}{99}
\bibitem[\protect\citeauthoryear{Chib \& Ramamurthy}{2010}]{chib} Chib, S., Ramamurthy, S. Tailored Randomized-block MCMC Methods with Application to DSGE Models, Journal of Econometrics, 155, 19-38 (2010)

\bibitem[\protect\citeauthoryear{Feroz, Hobson, 
\& Bridges}{2008}]{2008arXiv0809.3437F} Feroz F., Hobson M.~P., Bridges M., MultiNest: an efficient and robust Bayesian inference tool for cosmology and particle physics, arXiv:0809.3437 (2008)

\bibitem[\protect\citeauthoryear{Marinari 
\& Parisi}{1992}]{1992EL.....19..451M} Marinari E., Parisi G., Simulated Tempering: A New Monte Carlo Scheme, Europhysics Letters, 19, 451 (1992)

\bibitem[Mukherjee et al.(2006)]{2006ApJ...638L..51M} Mukherjee, P., 
Parkinson, D., Liddle, A.~R. A Nested Sampling Algorithm for 
Cosmological Model Selection. The Astrophysical Journal 638, L51-L54 (2006) 

\bibitem[\protect\citeauthoryear{Murray}{2007}]{murray} Murray, I., Advances in Markov chain Monte Carlo methods. PhD thesis, Gatsby computational neuroscience unit, University College London (2007)

\bibitem[\protect\citeauthoryear{Neal}{2003}]{slice} Neal, R. M., Slice sampling (with discussion), Annals of Statistics, vol. 31, pp. 705-767 (2003)

\bibitem[\protect\citeauthoryear{P{\'a}rtay, Bart{\'o}k, 
\& Cs{\'a}nyi}{2009}]{2009arXiv0906.3544P} P{\'a}rtay L.~B., Bart{\'o}k A.~P., Cs{\'a}nyi G., Full sampling of atomic configurational spaces, arXiv:0906.3544 (2009)

\bibitem[\protect\citeauthoryear{Roberts, Gelman \& Gilks}{1997}]{roberts} Roberts, G.~O., Gelman, A., Gilks, W.~R., Weak convergence and optimal scaling of random walk Metropolis algorithms, Annals of Applied Probability, Volume 7, Number 1 (1997), 110-120.

\bibitem[\protect\citeauthoryear{Rosenthal}{2010}]{rosenthal}
Rosenthal, J.~S. Optimal proposal distributions and adaptive MCMC. In S. P.
Brooks, A. Gelman, G. Jones, and X.-L. Meng (Eds.), Handbook of Markov Chain
Monte Carlo. Chapman and Hall/CRC Press (2010)

\bibitem[\protect\citeauthoryear{Sivia \& Skilling}{2006}]{sivia} Sivia, D.~ S., Skilling, J., Data Analysis: A Bayesian Tutorial, 2nd Edition,  Oxford University Press (2006)

\bibitem[\protect\citeauthoryear{Skilling}{2006}]{skilling} Skilling, J., Nested Sampling for General Bayesian Computation, Bayesian Analysis 4, pp. 833-860 (2006)

\bibitem[\protect\citeauthoryear{Trias, Vecchio \& Veitch}{2009}]{veitch} Trias, M., Vecchio, A. and Veitch, J. Delayed rejection schemes for efficient Markov-Chain Monte-Carlo sampling of multimodal distributions. arXiv:0904.2207 (2009)

\bibitem[\protect\citeauthoryear{Wang \& Landau}{2001}]{wl} Wang, F. and Landau, D.~P., Efficient, Multiple-Range Random Walk Algorithm to Calculate the Density of States, Phys Rev Letters 86, 2050 (2001)
\end{thebibliography}
\end{document}